\documentclass{kapproc} 

\kluwerbib

\begin{document}

\articletitle{Metal abundance of the eclipsing binary YZ Cas}

\articlesubtitle{Discrepancy between atmospheric determinations for the Am 
component, YZ Cas A}

\author{E. Lastennet}
\affil{Astronomy Unit, QMW, 
    Mile End Road, London E1 4NS, UK}

\author{D. Valls-Gabaud}
\affil{Lab. d'Astrophysique, UMR CNRS 5572, OMP, 
14 av. Belin, F-31400 Toulouse, France}

\author{C. Jordi}
\affil{Dep. d'Astronomia i Meteorologia, 
       Av. Diagonal 647, E-08028 Barcelona, Spain}


\begin{abstract}

We review current and new estimates of the eclipsing binary YZ Cassiopeiae
metallicity (Z).  Since the individual components cover a quite large range of
mass (1.35-2.31M$_{\odot}$), YZ Cas is potentially
one of the best stellar laboratories to understand the structure and evolution
of 1 to 2 M$_{\odot}$ stars.  
The derivation of Z from IUE spectra, as well as from photometric indices,
provides the chemical composition of the atmosphere (Z$_{\rm atmospheric}$), 
while the fit of evolutionary tracks provides the initial chemical composition 
(Z$_{\rm initial}$). 
While a disagreement is expected between Z$_{\rm atmospheric}$ and Z$_{\rm initial}$ 
because the primary component is an Am star (one expects Z$_{\rm atmospheric}$ to be 
larger), we find some unexpected discrepancy between atmospheric determinations of Z 
for this star. 

\end{abstract}


\section{Z$_{\rm atmospheric}$ from IUE spectra}

From IUE spectra of the YZ Cas primary, [LM83] found that the metals  
add up to a Z value of over 0.03 (Z$=$0.036$\pm$0.005).
We try to assess to what extent this high Z-value has to be revised. 
It is well known that the abundance of iron is enhanced in Am stars 
with respect to normal  A-F stars. 
[LM83] found for YZ Cas A that Fe is $\sim$6  times overabundant, 
which implies that Fe contributes to $\sim$43\% of Z. 
In order to update the [LM83] study,  
we applied two corrections to this value: 
1) 
They assumed an iron solar abundance A$_{Fe}$$=$7.60.  
More detailed studies give A$_{Fe}$ $=$ 7.50$\pm$0.05 ([GS99])
implying that Fe contributes to $\sim$57\% of Z.
2) Their analysis assumed T$_{\rm eff}$$=$10,300K. 
Recent determinations give 9100$\pm$300K ([R00]). 
This would reduce the [LM83]'s iron abundance by about 0.15-0.20 dex, 
and then imply that Fe contributes to $\sim$33\% of Z.  
Both corrections are opposite and imply altogether that Fe
contributes to $\sim$40\% of Z. 
Hence, our attempt to update some of the [LM83] assumptions 
does not change their high-Z conclusion.  

\section{Z$_{\rm atmospheric}$ from photometric calibrations}

Since $uvby\beta$ Str{\"o}mgren colours are available for both   
components of YZ Cas ([L81], [J97]), 
it is possible to derive [Fe/H] from existing calibrations.  
The [J97] photometric indices do not suggest a metallic
behaviour of the Am star atmosphere: [Fe/H] is $-$0.02$\pm$0.22 ([J00]), 
i.e Z$=$0.017$^{+0.012}_{-0.007}$ (assuming Z$_{\odot}$$=$0.018). 
The upper limit is only marginally compatible with the Z derived from the 
IUE spectra. 
Assuming the individual photometric indices from [L81], the
derived atmospheric [Fe/H] is 0.12$\pm$0.05, i.e. Z$=$0.024$\pm$0.003, 
again only marginally compatible with the IUE determination. 
Using Str{\"o}mgren photometry from [J97] and accurate gravities, [L99] 
derived simultaneous T$_{\rm eff}$-[Fe/H] estimates from the BaSeL 
photometric calibrations ([L98]).
They obtain [Fe/H] strictly lower than 0.1 for YZ Cas A, i.e Z$<$0.022.
This result assumes no reddening which is justified by experiments we
performed with the BaSeL models 
(see [L99] for details on the method) 
for E(b$-$y) ranging from $-$0.034 ({\it negative} value of [L81]) to 
0.09 (maximum value quoted by [LM83]).  
In summary, even if the results of [J00] and [L99]
are not completely independent because based on the same colour 
indices, they suggest that the atmospheric metallicity of YZ Cas A is solar 
or sub-solar, and exclude Z-values greater than 0.03.

\end{document}